%% file: _main.tex
\documentclass[preprint,journal]{vgtc}       % preprint (journal style)

\ifpdf%                                % if we use pdflatex
  \pdfoutput=1\relax                   % create PDFs from pdfLaTeX
  \usepackage{graphicx}                % allow us to embed graphics files
  \DeclareGraphicsExtensions{.pdf,.png,.jpg,.jpeg} % for pdflatex we expect .pdf, .png, or .jpg files
\else%                                 % else we use pure latex
  \usepackage{graphicx}                % allow us to embed graphics files
  \DeclareGraphicsExtensions{.eps}     % for pure latex we expect eps files
\fi%

\graphicspath{{figures/}{pictures/}{images/}{./}} % where to search for the images
\PassOptionsToPackage{bookmarks=false}{hyperref}% to pass the IEEE
\usepackage{microtype}                 % use micro-typography (slightly more compact, better to read)
\PassOptionsToPackage{warn}{textcomp}  % to address font issues with \textrightarrow
\usepackage{textcomp}                  % use better special symbols
\usepackage{mathptmx}                  % use matching math font
\usepackage{times}                     % we use Times as the main font
\usepackage{cite}                      % needed to automatically sort the references
\usepackage{tabu}                      % only used for the table example
\usepackage{booktabs}                  % only used for the table example

\PassOptionsToPackage{hyphens}{url} % make sure overly long URLs are properly hyphenated

\usepackage{microtype}

\usepackage{algpseudocode, caption}
\DeclareCaptionType{algorithm}

\input{exsitu}

\usepackage{hyperref}
\usepackage{nameref}

\usepackage{sepfootnotes}

\usepackage[para]{footmisc}

\usepackage{xspace} 

\usepackage[normalem]{ulem}

\usepackage{float}
\floatstyle{boxed}
\newfloat{floatingtextbox}{thp}{lop}
\floatname{floatingtextbox}{Box}

\usepackage[title]{appendix}

\newif\ifanonymous \anonymousfalse 
\newif\ifdraft \draftfalse
\newif\ifcomments \commentsfalse
\newif\ifcameraready \camerareadyfalse
\newif\ifsubfigures \subfigurestrue
\newif\ifligature \ligaturetrue

\draftfalse
\ifligature
\else
\usepackage{microtype}
\DisableLigatures{encoding = *, family = *}
\fi

\ifdraft
\newcommand{\alex}[1]{\textcolor{RoyalBlue}{#1}}
\newcommand{\xiaoyi}[1]{\textcolor{Purple}{xiaoyi: #1}}
\newcommand{\chat}[1]{\textcolor{Green}{chat: #1}}
\newcommand{\wem}[1]{\textcolor{Teal}{wem: #1}}

\newcommand{\remove}[1]{\textcolor{Red}{\sout{#1}}}

\newcommand{\changed}[1]{\textcolor{RoyalBlue}{#1}}

\else
\newcommand{\alex}[1]{}
\newcommand{\chat}[1]{}
\newcommand{\xiaoyi}[1]{}
\newcommand{\wem}[1]{}

\newcommand{\remove}[1]{}

\newcommand{\changed}[1]{#1}

\fi

\newcommand{\argus}[0]{\textit{Argus}\xspace}
\newcommand{\Argus}[0]{\argus}

\newcommand{\viewConfound}[0]{\textit{Confound} sliders\xspace}
\newcommand{\viewTradeOff}[0]{\textit{Power Trade-off} view\xspace}
\newcommand{\viewHistory}[0]{\textit{History} view\xspace}
\newcommand{\viewExpectedAverages}[0]{\textit{Expected-averages} view\xspace}
\newcommand{\viewPairwise}[0]{\textit{Pairwise-difference} view\xspace}
\newcommand{\viewExperimentDesign}[0]{\textit{Experiment-design} view\xspace}

\newcommand{\ViewTradeOff}[0]{\textit{Power Trade-off} View\xspace}
\newcommand{\ViewHistory}[0]{\textit{History} View\xspace}
\newcommand{\ViewExpectedAverages}[0]{\textit{Expected-averages} View\xspace}
\newcommand{\ViewPairwise}[0]{\textit{Pairwise-difference} View\xspace}
\newcommand{\ViewExperimentDesign}[0]{\textit{Experiment-design} View\xspace}

\newcommand{\pSeven}[0]{P7\textsubscript{N}\xspace}
\newcommand{\pEight}[0]{P8\textsubscript{N}\xspace}
\newcommand{\pNine}[0]{P9\textsubscript{N}\xspace}

\newcommand{\ReadingTime}[0]{\textsc{ReadingTime}\xspace}
\newcommand{\Medium}[0]{\textsc{Medium}\xspace}
\newcommand{\Paper}[0]{\textsc{Paper}\xspace}
\newcommand{\Screen}[0]{\textsc{Screen}\xspace}
\newcommand{\Layout}[0]{\textsc{Layout}\xspace}
\newcommand{\OneColumn}[0]{\textsc{One\_Column}\xspace}
\newcommand{\TwoColumn}[0]{\textsc{Two\_Column}\xspace}

\newcommand{\cLinear}[0]{\textsc{Linear}\xspace}
\newcommand{\cDesigner}[0]{\textsc{Designer}\xspace}
\newcommand{\cBayesian}[0]{\textsc{Bayesian}\xspace}
\newcommand{\cKMeans}[0]{\textsc{K-Means}\xspace}

\newcommand{\aCohenD}[0]{Appendix A\xspace}
\newcommand{\aPropagation}[0]{Appendix B\xspace}
\newcommand{\aArchitecture}[0]{Appendix C\xspace}
\newcommand{\aThinkAloud}[0]{Appendix D\xspace}

\graphicspath {{_figures/}}

\ifanonymous
\else

\fi

\anonymousfalse
\onlineid{1070}

\vgtccategory{Research}
\vgtcpapertype{application/design study}

\title{Argus: Interactive \textit{a priori} Power Analysis}

\author{Xiaoyi Wang, Alexander Eiselmayer, Wendy E. Mackay, Kasper Hornbæk, Chat Wacharamanotham}
\authorfooter{
	\item Xiaoyi Wang and Kasper Hornbæk are with the University of Copenhagen, Denmark. E-Mail: \{xiaoyi.wang, kash\}@diku.dk.
	\item Alexander Eiselmayer and Chat Wacharamanotham are with the University of Zurich, Switzerland. E-Mail: \{eiselmayer, chat\}@ifi.uzh.ch.
	\item Wendy E. Mackay is with Univ. Paris-Sud, CNRS, Inria, Universit\'e Paris-Saclay, France. E-Mail: mackay@lri.fr.
}

\shortauthortitle{Wang \MakeLowercase{\textit{et al.}}: Argus: Interactive a priori Power Analysis}
\abstract{
A key challenge HCI researchers face when designing 
a controlled experiment is choosing the appropriate number of participants, or sample size.
\textit{A priori} power analysis examines the relationships among multiple parameters, including the complexity associated with human participants, \eg, order and fatigue effects, 
to calculate the statistical power of a given experiment design.
We created \argus, a tool that supports interactive exploration of statistical power:
Researchers specify experiment design scenarios with varying confounds and effect sizes.
\argus then simulates data and visualizes statistical power across these scenarios, 
which lets researchers interactively weigh various trade-offs and make informed decisions about sample size. 
We describe the design and implementation of \argus, %
\changed{%
 a usage scenario designing a visualization experiment, and a think-aloud study.
 }%

} % end of abstract

\keywords{Experiment design, power analysis, simulation}

\CCScatlist{ % not used in journal version
 \CCScat{K.6.1}{Management of Computing and Information Systems}%
{Project and People Management}{Life Cycle};
 \CCScat{K.7.m}{The Computing Profession}{Miscellaneous}{Ethics}
}

\teaser{
  \centering
  \includegraphics[width=\linewidth]{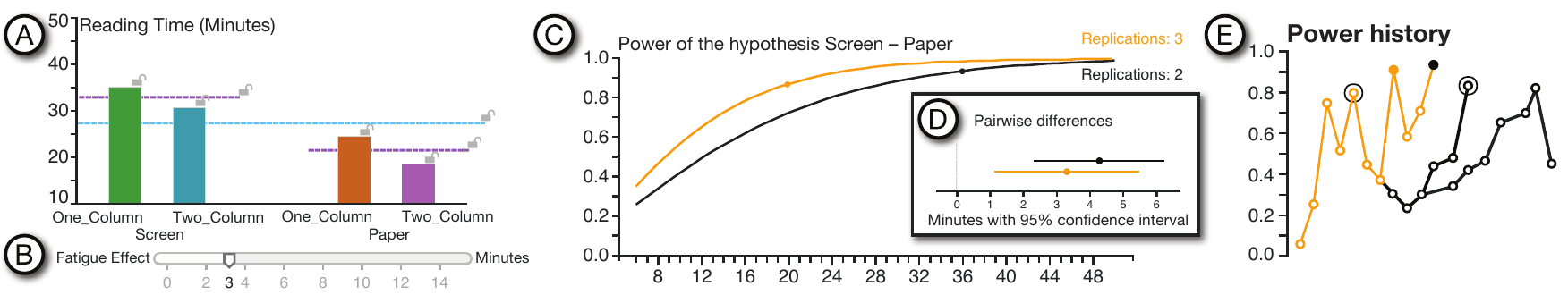}
  \caption{Argus interface: 
  \changed{(A) Expected-averages view helps users estimate the means of the dependent variables through interactive chart. 
  (B) Confound sliders incorporate potential confounds, \eg, fatigue or practice effects.
  (C) Power trade-off view simulates data to calculate statistical power; and
  (D) Pairwise-difference view displays confidence intervals for mean differences, 
  animated as a \textit{dance of intervals}.
  (E) History view displays an interactive power history tree so users can quickly compare statistical power with previously explored configurations.}
}
  \label{fig:teaser}
}

\renewcommand{\manuscriptnotetxt}{}

\vgtcinsertpkg

\begin{document}


%\firstsection{Introduction}

\maketitle
\input{_sections/01_introduction}
\input{_sections/02_background_task_analysis}

\input{_sections/03_related_work}

\input{_sections/06_argus}

\input{_sections/06-1_implementation_details}
\input{_sections/06-2_use_case}
\input{_sections/07_think-aloud_study}
\input{_sections/09_lesson_learned.tex}
\input{_sections/08_Discussion.tex}
\input{_sections/99_conclusion}

% \input{_sections/xx_power_analysis_at_CHI.tex}

\acknowledgments{
This work is partially is supported by the Innovation Fund Denmark, the BIOPRO2 strategic research center grant № 4105-00020B, the European Research Council (ERC) grants № 695464 ``ONE: Unified Principles of Interaction'', and the University of Zurich GRC Travel Grant. We also thank Michel Beaudouin-Lafon for initial feedback and some vision directions in the beginning of the project.
%The authors wish to thank A, B, and C. This work was supported in part by
%a grant from XYZ (\# 12345-67890).
%Left blank for review.
% we should acknowledge Michel for initial feedback and some vision directions in the beginning
% we should add Michel's ERC ONE grant so that Alex was able to visit Paris for the prelimanry study
}

\bibliographystyle{abbrv-doi}

\bibliography{_references}
\balance{}

\end{document}

%% file: exsitu.tex
\usepackage{balance}  % to better equalize the last page
\usepackage{graphics} % for EPS, load graphicx instead 
\usepackage[T1]{fontenc}
\usepackage{mathptmx}
\usepackage[pdftex]{hyperref}
\usepackage[usenames,dvipsnames,svgnames]{xcolor}
\usepackage{booktabs}
\usepackage{textcomp}
\usepackage{microtype} % Improved Tracking and Kerning
\usepackage{ccicons}  % Cite your images correctly!
\usepackage[utf8]{inputenc} % for a UTF8 editor only

\usepackage{csquotes}
\MakeOuterQuote{"}

\makeatletter
\def\url@leostyle{%
  \@ifundefined{selectfont}{
    \def\UrlFont{\sf}
  }{
    \def\UrlFont{\small\bf\ttfamily}
  }}
\makeatother
\def\pprw{8.5in}
\def\pprh{11in}

\definecolor{linkColor}{RGB}{6,125,233}
\usepackage{MnSymbol} % for math symbols
\usepackage{enumitem} % to control spacing in enumerations
\usepackage{multirow} % to support tables with cells spanning multiple rows
\usepackage{xspace} % to avoid having to type {} for macros with no arguments

\makeatletter%
\enitkv@key{enumitem}{nosep}[true]{%
  \parskip=\z@skip
  \partopsep=\z@skip
  \topsep=\z@skip
  \itemsep=\z@skip
  \parsep=\z@skip}
\enitkv@key{enumitem}{notopsep}[true]{%
  \parskip=\z@skip
  \partopsep=\z@skip
  \topsep=\z@skip
  \itemsep=\z@skip
  \parsep=\z@skip}
\makeatother%

\newcommand\hfigurerow[3]{\let\subfigitem\fcapside\newcommand\emptycaption{\caption{}}%
\ffigbox{}{\CommonHeightRow{\begin{subfloatrow}[#1]#2\end{subfloatrow}}{#3}}}
\newcommand\vfigurerow[3]{\let\subfigitem\ffigbox\newcommand\emptycaption{}%
\ffigbox{}{\CommonHeightRow{\begin{subfloatrow}[#1]#2\end{subfloatrow}}{#3}}}
\newcommand\defineComment[3]{%
\ifcomments%
\expandafter\newcommand\csname #1\endcsname[1]{{\leavevmode\color#2{#3##1}}}%
\else%
\expandafter\newcommand\csname #1\endcsname[1]{\ignorespaces}%
\fi}

\newcommand\definePersistentComment[3]{%
\ifcomments%
\expandafter\newcommand\csname #1\endcsname[1]{{\leavevmode\color#2{#3##1}}}%
\else%
\expandafter\newcommand\csname #1\endcsname[1]{##1}%
\fi}

\def\firstof#1 #2\nil{#1}
\def\restof#1 #2\nil{#2}

\newcommand\authorColors{BlueViolet FireBrick SlateBlue RedOrange TealBlue WildStrawberry RoyalPurple SteelBlue}
\newcommand\nextAuthorColor{\expandafter\firstof\authorColors\nil}
\newcommand\popAuthorColor{\edef\authorColors{\expandafter\restof\authorColors\nil}}

\def\expandArg #1#2{\expandafter\expandArgaux\expandafter{#2}{#1}}
\def\expandArgaux #1#2{#2{#1}}

\makeatletter
\newcommand\defineAuthorComment[2][]{%
\def\auth@name{#1}\ifx\auth@name\empty\def\auth@name{#2}\fi% get the macro name (defaults to author name)
\expandafter\edef\csname color@\auth@name\endcsname{\nextAuthorColor}% create a macro whose name is color@<macroname> that holds the color of that user
\expandArg{\defineComment{\auth@name}}{\csname color@\auth@name\endcsname}{#2:}% call defineComment
\popAuthorColor% pop the first color so that next call uses a different color
}
\makeatother

\newcommand*{\eg}{e.g.\@\xspace}
\newcommand*{\ie}{i.e.\@\xspace}
\newcommand*{\vs}{vs.\@\xspace}

\makeatletter
\newcommand*{\etal}{%
    \@ifnextchar{.}%
        {et~al}%
        {et~al.\@\xspace}%
}
\newcommand*{\etc}{%
    \@ifnextchar{.}%
        {etc}%
        {etc.\@\xspace}%
}
\makeatother

\makeatletter
\newcommand\new@expname[4]{\def\exp@cmd{#1}\ifx\exp@cmd\empty\def\exp@cmd{#2}\fi\expandafter\newcommand\csname #3\exp@cmd\endcsname[0]{#4\xspace}}
\newcommand\newFactor[2][]{\new@expname{#1}{#2}{f}{{{\sc #2}}}}
\newcommand\newValue[2][]{\new@expname{#1}{#2}{v}{{{\it #2}}}}
\newcommand\newMeasure[2][]{\new@expname{#1}{#2}{m}{{\small {\it \MakeUppercase{#2}}}}}
\makeatother

%% file: _sections/01_introduction.tex
%!TEX root=../_main.tex

% --------------------------------------------------------------------
\section{Introduction}
% --------------------------------------------------------------------
Determining \emph{sample size} is a major %crucial 
challenge when designing  experiments with human
participants, \eg, in Information Visualization (VIS) and Human-Computer Interaction (HCI)~\cite{Eiselmayer2019, Hornbaek2013,  Lazar2017}.
% WEM Spell out acronyms the first time. 
%
% WEM I want to say this in a more positive way, 
% to begin with the benefits of estimating statistical power
Researchers want to save time and resources by choosing
the minimum number of participants that let them reliably detect an effect that truly exists in the population. 
However, if they underestimate the sample size, 
\ie the experiment lacks statistical power,
they risk missing the effect -- a Type II error.
% They are also less likely to publish these negative or null results, the so-called 
Researchers are also less likely to publish these negative or null results, the so-called 
``file drawer problem''~\cite{Rosenthal1979}.
Researchers cannot simply add participants until the results are significant, which is considered a malpractice,
and are strongly encouraged to preregister the sample size 
to increase the credibility of the investigation~\cite{Cockburn2018}.

% WEM -- old version
% Underestimating the sample size can lead to inadequate statistical power and failures to discover an effect 
% despite its presence in the population, a Type II error.
% This would contribute to negative or null results not being published, the file drawer problem~\cite{Rosenthal1979}.
% Because adding participants until the result is significant 
% is malpractice,
% preregistering the sample size increases the credibility of the investigation~\cite{Cockburn2018}. 

%A large sample size could be obtained by increasing the number of participants or increasing the number of replications. 
%The former could be is uneconomical due to the cost of participant recruitment, and the latter could lengthen the experiment, introducing a confounding fatigue effect. 
%Lastly, keep adding participants until the result is significant is a malpractice~\cite{Cockburn2018}. 
The sample size can be determined statistically with an \textit{a priori} power analysis.
However, %such an analysis 
this requires approximating the \emph{effect size}, which quantifies the strength and consistency of the influences of the experimental conditions on the measure of interest.
Estimating an effect size 
must account for
% is difficult because of 
the relationships between experimental conditions; 
the inherent variability of the measures, \eg, differences among study participants;
and variation in the structure of the experiment conditions, \eg, blocking and order effects. 
%The latter are the consequences of the choices in counterbalancing designs.
% The difficulty in 
This complexity acts as
% estimating effect size is
a major barrier to performing power analysis~\cite{Lipsey2009,Murphy2014}.
% Lipsey1990, p.47; Murphy2014, p. 17
\newpage % This page break simulates the footer that will appear in the camera-ready version. Do not remove
Studies in the natural sciences can rely on meta-analyses of multiple replication studies to suggest effect and sample sizes.
However, in VIS and HCI, such replications are rare~\cite{Hornbaek2007, Kosara2018} and not highly valued~\cite{Greenberg2008}. 
Sample sizes ($N$) are often chosen based on rules of thumb
% Therefore, sample sizes could be decided based on a rule of thumb (
\eg, $N$ $\geq$ 12~\cite{Eiselmayer2019},
% only a small number of studies 
or drawn from small numbers of studies~\cite{Caine2016, Hwang2010, Hornbaek2007}.
Studies with human participants also risk
% In studies with human participants, the structure of the experimental conditions could cause
\emph{confounding effects} such as fatigue, carry-over, and learning effects.
Analytical methods implemented with power analysis tools such as \texttt{pwr}~\cite{Champely2018} or G*Power~\cite{Faul2007}, are not usually sophisticated enough to account for these effects.
Furthermore, researchers must often weigh the benefit of statistical power against 
high recruitment costs, overly long experiment duration, 
and the inconvenience % WEM parallel structure
% recruitment costs, experiment duration, and convenience 
of switching between experiment conditions~\cite{Mackay2007}. % removed TS2
Although several interactive tools help
% Multiple interactive tools have been proposed to allow researchers to 
explore trade-offs among plausible experiment design configurations~\cite{Eiselmayer2019, Mackay2007, Meng2017}, 
few address the complex
% Yet, the complicated 
relationship between statistical power and relevant experiment parameters. %~\cite{Eiselmayer2019}.
% has only been touched upon~\cite{Eiselmayer2019}.
%
% WEM: Removed 2 TS2 references.
%
% \alex{Is it ok to cite TS2 three times in the last two sentences?}

%Incorporating these effects in power analysis is non-trivial, and usually could not be done analytically.
%Researchers usually attempt control these extraneous effects by deploying various counterbalancing techniques.
%However, complex experiment designs lead to difficulties in power analysis because .

Existing power analysis tools are designed as calculators: 
The user specifies acceptable Type I and Type II error rates, test statistics, experimental design, and an approximate size of the effect.
The tool then produces either a single sample size or a chart showing how statistical power increases in conjunction with the sample size, at several effect sizes.
We argue that researchers need tools for exploring
possible trade-offs between statistical power and the costs of other experimental parameters, especially when the effect size is uncertain.

\changed{%
We propose 
\argus, an interactive tool 
for exploring the relationship 
between sample size and statistical power,
given particular configurations of the experimental design.
Users can  
estimate parameters -- effect sizes, confounding effects, the number of replications, and the number of participants -- and see how they influence statistical power and the likely results in an interactive data simulation.
}

\noindent
\changed{%
\textbf{Contributions:}
We identify challenges and analyze the tasks involved in \textit{a priori} power analysis. We propose \argus---which combines interactive visualization and simulation to aid exploration and decision-making in experiment design and power analysis.
To demonstrate its efficacy, we describe a use case and a think-aloud study.
}
% This paper
% begins with a 
% task analysis of \textit{a priori} power analysis,
% followed by a discussion of related work.
% We then describe
% \argus, an interactive tool 
% for exploring the relationship 
% between sample size and statistical power,
% given particular configurations of the experimental design.
% Users can  
% estimate parameters -- effect sizes, confounding effects, the number of replications, and the number of participants -- and see how they influence statistical power and the likely results in an interactive data simulation.
% We describe the results of  
% a preregistered observational think-aloud study with nine novice and experienced 
% VIS and HCI researchers 
% to validate the potential of \argus 
% for improving the quality of experiment designs.
% We conclude with a discussion of directions for future research.

%% file: _sections/02_background_task_analysis.tex
%!TEX root=../_main.tex

% --------------------------------------------------------------------
\section{Background and Task Analysis}
% --------------------------------------------------------------------

\sepfootnotecontent{a_priori}%
{Although one can calculate achieved power from data collected during an experiment, such post-hoc analysis is impractical for planning experiments
or interpreting the results~\cite[p. 110]{Cairns2019} and \cite[section 5.9.4]{Yatani2016}. 
This paper thus uses the term `power analysis' to refer to \textit{a priori} power analysis.}

When planning an experiment, researchers use a strategy called \textbf{\textit{a priori} power analysis\sepfootnote{a_priori}} to choose which sample size will 
allow the experiment to detect an expected effect.
Power analysis uses the relationship between the \textbf{sample size} and the following parameters:

\begin{itemize}[itemindent=-4mm, leftmargin=4mm, itemsep=0pt] 

\item[] \textbf{$\pmb{\alpha}$} is the probability of detecting an effect from an experiment when it is actually absent in the population (Type I error: false alarm).
Researchers usually set $\alpha$ based on the convention of each academic field, typically .05 for VIS, 
HCI, psychology, and the social sciences. 

\item[] \textbf{$\pmb{1 - \beta}$}, or statistical power, 
is the probability that a long run of experiments will successfully 
detect an effect that is true in the population.
($\beta$ is the probability of a Type II Error: missing the true effect.) 
If no existing basis exists, Cohen proposed a convention of 0.8~\cite[p.56]{Cohen1988}.

\item[] \textbf{Effect size} is the difference %
across means calculated from data under each condition. 
Researchers make an educated guess of the effect size based on previous research or their experience.
Effect sizes are standardized for the calculation, as described in C3 below. 
\end{itemize}

The sample size can be calculated with these parameters, either with software or from a statistics textbook \eg~\cite{Cohen1988}.
When facing resource constraints, such as personpower, time or budget,
researchers sometimes sacrifice statistical power in exchange for a more attainable sample size.
In cases where access to participants is limited \eg patients, children or other special populations, 
power analysis may be skipped altogether.
Even if
the power analysis suggests an unrealistic sample size, it might still offer a useful cost-benefit assessment.
In any case, researchers who choose to conduct a power analysis still face the following challenges:
\vspace{0.2em}

\textbf{C1: Estimating a reasonable effect size is difficult.}
Researchers who wish to estimate the effect size face a paradox:
The goal of conducting the experiment is to discover the true effect size 
in the population, but selecting the correct sample size for revealing that effect requires an initial estimate of the effect size.
Overestimating the effect size often leads to a sample size that exceeds available resources.
Even for studies that can easily scale up the sample size, using an overly
large sample size is ``wasteful'' and an ``unethical'' use of study participants’ time~\cite{Button2013}.
Although researchers can conduct pilot studies,
finding a large effect size in a pilot with few participants may be misleading and 
result in an underpowered final experiment~\cite[p. 280]{Lakens2014b}.
Cohen proposed a guideline for standardized effect sizes 
derived from data on human heights and intelligence quotients~\cite{Cohen1977}.
However, reviews in 
domains such as software engineering~\cite{Kampenes2007} found that the distribution of effect sizes from experiments differ from Cohen’s guideline.
Therefore, many researchers recommend against using guidelines that are not specific to the domain of study~\cite{Lenth2001, Cummings2011, Baguley2004}.
In fields where replication studies are scarce, \eg, 
VIS and HCI~\cite{Kosara2018, Hornbaek2014}), 
researchers must generate possible effect-size scenarios.

%\noindent
%\textit{Design implication: } Users should be able to easily specify and interpret effect size.
%\vspace{0.2em}

\begin{figure} %[htb]
\centering
  \includegraphics{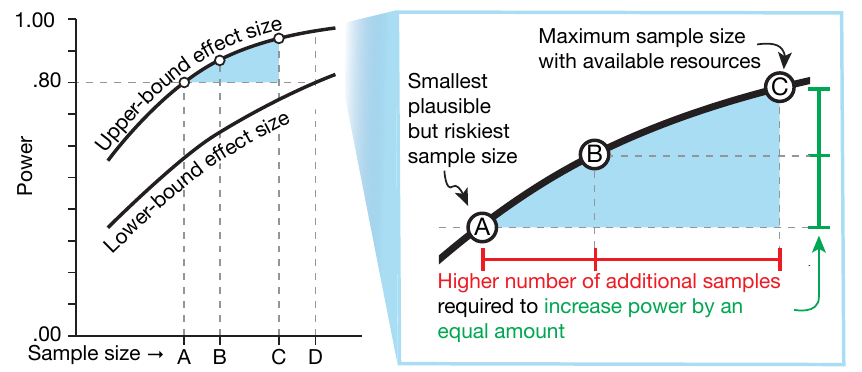} %[width=0.5\columnwidth]
  \caption{Determining power and sample size with effect-size uncertainty and resource constraints.
  }
    \label{fig:power_decision}
\end{figure}

\textbf{C2: Comparing power at multiple effect size scenarios is necessary.}
Instead of estimating a single value for the effect size, 
some researchers estimate the upper-bound––to represent the best case––and the lower-bound––below which the effect is too small to be practically meaningful~\cite[p. 57]{Lenth2001, Lipsey2009}––which results in a range of sample sizes to consider 
(\autoref{fig:power_decision}, A--D).
However, in many experiments, the largest attainable sample size may be lower than the one required by the lower-bound effect size (\autoref{fig:power_decision}, C).
Researchers must then weigh the benefit of further mitigating risk by increasing the power and the cost of a larger sample size.
Because the function between the power and sample size is concave, 
improving power is increasingly costly~\cite[p. 702]{Lakens2014a} (\autoref{fig:power_decision}, A--B vs. B--C).
Among existing software for calculating statistical power, only a few plot the statistical power and the sample size at different effect sizes 
(see Related Work).

%\noindent
%\textit{Design implication: } Users should be able to compare power at multiple effect-size estimations.

%\sepfootnotecontent{operative}%
%{Lipsey \& Hurley \cite[p. 53–61]{Lipsey2009} further distinguish between the widely-used standardized effect sizes (e.g., Cohen’s d) and operative effect sizes of which standardizer is further adjusted specifically to the experiment design.}

\newpage
\textbf{C3: Standardized effect sizes are not intuitive.} 
The difference between means is an example of a \emph{simple effect size}, which is based on the original unit of the dependent variable and thus has intuitive meaning for researchers.
However, power calculation requires a \emph{standardized effect size}, which is calculated by dividing the simple effect size with a standardizer.
The formula for the standardized effect size depends on how the sources of the variances are structured, which in turn depends on the experiment design.
\changed{%
(See \aCohenD for an example on how blocking influences calculation of effect size.) %
}
Note how an estimate in the form of a simple effect size may yield different standardized effect sizes. 
Researchers often have difficulty using standardized effect sizes when choosing their sample size, since these are ``not meaningful to non-statisticians''~\cite{Baguley2004}.

%\noindent
%\textit{Design implication:} Users should be able to work with intuitive simple effect sizes, rather than non-intuitive standardized effect sizes.

\changed{%
\textbf{C4: Power analysis excludes the temporal aspect of experiment design.}
}
Power analysis simplifies sources of variations into a few standard deviations within effect size formul\ae. 
(See \aCohenD for an example.) 
Potential confounds---\eg, the fatigue effect or the practice effect---lose their temporality once encoded into standard deviations. 
This loss could be a reason that separates power analysis from the rest of the experiment design process~\cite{Eiselmayer2019}. 
Better integration of temporal effects and design parameters---\eg, number of replications and how conditions are presented to study participants---could allow better exploration of trade-offs.

%\noindent
%\textit{Design implications:} Users should be able to explore power trade-off in light of temporal effects and design parameters.

%----------------------------------------------------
\subsection{\changed{Task Analysis}}

Under the What-Why-How framework~\cite{Brehmer2013, Munzner2015}, the task abstraction could be described as follows. 
All of the attributes below are quantitative unless stated otherwise.

\textbf{T1: Come up with an effect size estimate.} Simple effect sizes—the difference in the responses between conditions—could have been estimated directly. 
Alternatively, the estimation can be simplified by first estimating the mean in a baseline experimental condition, and then \texttt{deriving} the value of other conditions by \texttt{comparing} each with the baseline. 
The conversion from the simple effect size to the standardized effect size (C3) could be automated when the information about experiment design is available in a computable form. 

\textbf{T2: Check the potential outcome effect size.}
For experiments with two independent variables or more, the possibilities of the interaction effects could obfuscate how the \textit{a priori} effect sizes influence the final results. 
(More details in~\autoref{sec:average}.) 
A data simulation could allow the users to \texttt{compare} the simulated effect sizes among themselves or \changed{to compare them with the specified input}---especially in the presence of interaction effects.

\textbf{T3: Determine candidate sample sizes.}
Researchers \texttt{browse} for the sample size with a reasonable trade-off within a set of constraints (e.g., resources for participant recruitment). 
To facilitate efficient browsing, they \texttt{identify} features of the relationship between power and sample sizes, \eg, where the power-gain is steep or where it plateaus. 
Multiple scenarios (C2) of effect sizes could also generate different relationships, leading to the need to \texttt{compare} their trends.

\textbf{T4: Try out potential scenarios.}
Due to uncertainties in effect size estimation (C1), researchers need to be able to \texttt{explore} the dependency between their effect size estimates and other parameters—\eg, the fatigue effect (C4)—to the power-sample size relationship. 
Thus, they need to be able to \texttt{record} and review the scenarios. 
Some changes to the scenarios are categorical—\eg, different choices of counterbalancing strategies. 
Others are quantitative—\eg, different amounts of the fatigue effect. 
The abstract data type of the scenarios could be a \emph{multidimensional table} with each input parameter as a key and the resulting power as an attribute. 
However, this abstraction does not capture researchers’ exploration traces. 
Such traces could be abstracted as a \emph{tree} in which each child node is a scenario that is derived based on its parent node.

%% file: _sections/03_related_work.tex
%!TEX root=../_main.tex

% --------------------------------------------------------------------
\section{Related Work} \label{sec:related_work}
% --------------------------------------------------------------------
% We first describe how existing software packages support the challenges in power analysis.
% Then, we describe findings and theories from visualization research to highlight the aspects that are lacking in these software packages.

% \subsection{Existing software tools for power analysis}

Before the prevalence of personal computers, % looked up the
researchers used
% \wm{looked up tables~\cite[pp. 28—39]{Cohen1988}) and charts~\cite{Scheffe1959}) in textbooks
look-up tables~\cite[pp. 28—39]{Cohen1988}) and charts~\cite{Scheffe1959}) in textbooks
to determine
the relationship between sample size, effect size, statistical power, and Type I error rate, usually fixed at .05.
% the tables (e.g.,~\cite[pp. 28—39]{Cohen1988}) or charts (e.g.,~\cite{Scheffe1959}) of textbooks.
% To alleviate the burden of this lookup,   ... provided
Early software packages simplified the process by providing
command-line or menu interfaces to specify parameters, and 
displayed
% yielded results as 
a single value for statistical power.
Goldstein~\cite{Goldstein1989} surveyed 13 power analysis software packages and 
highlighted %pointed out 
the lack of two key functions: % functionalities for statistical power calculation: 
plotting a chart of the trade-offs between parameters, and 
capturing intermediate results for comparison.
Borenstein et al.~\cite{Borenstein1992} pioneered the use of visualization to specify input parameters and inspect relationships among parameters.
For input, the tool shows a box plot of the dependent variable by condition on the screen.
The simple effect size can be specified by moving the mean and standard deviation of each group with arrow keys or function keys.
The software then outputs the effect size and power in real-time.
It also produces a chart showing the relationship between power and sample size under multiple effect-size scenarios (see \autoref{fig:power_decision}, left).
Nevertheless, due to the low screen resolution, the relationship chart is presented on a separate screen from the input specification, hindering interactive exploration.
This tool also 
restricts analysis to
% only supports the analysis of 
between-subjects designs with two conditions
and does not support  % No 
exploration of the impact of choices in experimental design. % is possible.

% At present, 
G*Power~\cite{Faul2004, Erdfelder1996, Faul2007} is one of the most widely used power analysis software tools today.
% because it is free and is available for both Windows and Mac.
G*Power developers prioritize covering multiple types of statistical tests and high-precision calculation rather then facilitating exploration~\cite{Erdfelder1996}.
% and have abandoned a graphical interface for in exchange for sophisticated
% they have abandoned a visual approach 
% parameter exploration~\cite{Erdfelder1996}.
%
% Chat's version
G*Power calculates power from one set of input parameters at a time.
%
%% Wendy's version
% Users specify input parameters with a menu interface that calculates
% a single output at a time.
%
% Alex's version
% Users specify input parameters with a menu interface from which
% a single output is calculated at a time.
%
% Users specify input parameters with a menu interface 
% % \wm{that calculates}  % and calculate 
% \alex{from which}
% a single output \alex{is calculated} at a time. \wem{Why convert to passive voice?} \alex{For me, it would read like the menu interface does the calculation rather then the tool itself. If active is nicer, I am absolutely happy to revert back :) }
%
This forces them to record % Then, they need to note down 
parameters and output at each step of the exploration process. % their exploration
% G*Power generates a static chart from a given range (from, to, step) standardized effect size.
G*Power generates a static chart from a given range of standardized effect sizes.

Some software packages integrate power analysis with experiment design.
JMP's design of experiment (DOE) function~\cite{JMPDOE} provides a menu interface for power calculation and generates static charts 
similar to those of G*Power.
% in the same manner as G*Power.
The R package \texttt{skpr}~\cite{Morgan-Wall2018} provides a menu-based interface for generating experiment designs.
However, it only calculates and shows a single power estimate at a time.
To explore different effect size scenarios, users must manually save and restore states via their web browser's bookmark function.
\texttt{skpr} provides a menu interface for generating experiment trial tables and calculating power. 
However, it provides only the power of the entire experiment design: all variables that take part in the counterbalancing contributes to the power analysis.
Touchstone2~\cite{Eiselmayer2019} provides a direct manipulation interface for specifying experiment design and displays an interactive chart that visualizes the relationship between the number of participants and power.
Unlike \texttt{skpr}, users can select a subset of independent variables to include in the power calculation.
This lets researchers include nuisance % This functionality allows nuance 
variables % to be included 
in the counterbalancing design, without affecting power calculation.
Even so, Touchstone2 does not include confounding effects 
and relies on menus to specify effect size.
% in the exploration and still uses a menu interface for effect size specification.

% \subsection{Visual elicitation and exploration for understanding}
Several researchers have shown that graphical user interfaces (GUI) are better than menus for specifying estimations.
% Several works demonstrate that graphical elicitation is superior to menu interfaces.
% Alex: Love what Wendy wrote, much simpler.
Goldstein \& Rothschild~\cite{Goldstein2014} compared numerical and  graphical interfaces % method 
to elicit laypeople's intuitions about the probability distributions of events. % Not clear -- check that this is what you mean
They show that users achieve greater accuracy 
when they can specify distributions graphically.
% Their results show that greater accuracy is achieved by letting users specify the distribution through graphical representation.
Hullman \etal~\cite{Hullman2018} 
support these results 
% confirm this effect   WEM Too strong ...
in the context of estimating effect sizes for experiments. % before seeing  results from replication experiments.  %% ??? WEM Don't understand
We argue that power analysis software would benefit from such graphical representations of  relationships among parameters, with a GUI to manipulate them.% This is more positive and lead better to the next section.

%% file: _sections/06_argus.tex
\begin{figure*}[ht!]
\centering
  \includegraphics[height=9cm]{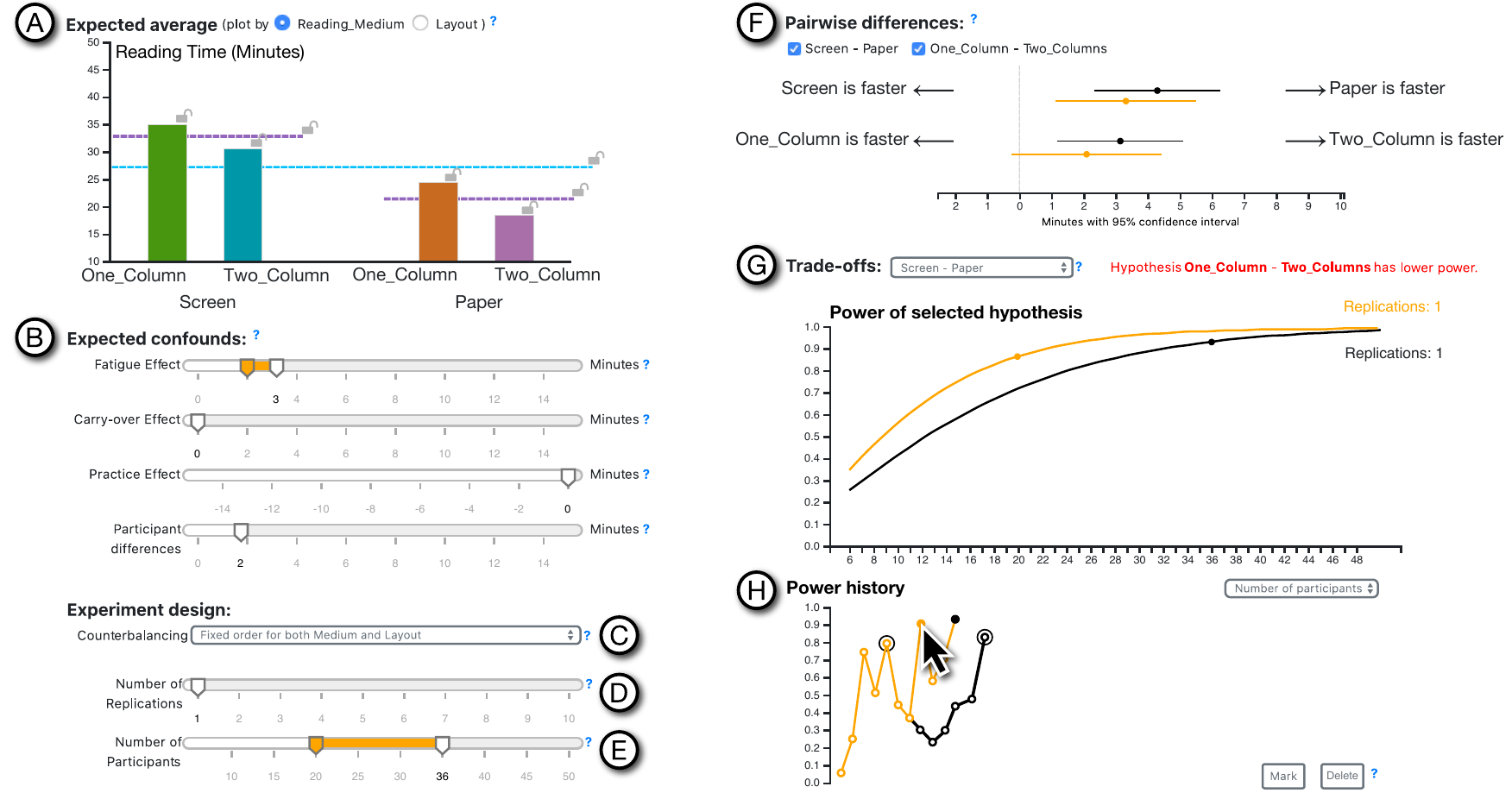}
  \caption{
  \argus interface: % Argus user interface.
  (Left:) Users estimate effect size by specifying: 
  (A) the expected % their expectation of 
  average for each condition;
%   They also specify the various
  (B) the relevant confounding effects, and 
  (C--E) the experimental design elements.
  (Right:) The simulation output includes: 
  (F) pairwise differences, with expected results %  in terms of the 
  shown as differences between means;
  (G) the relationship between power and sample size % shown in (G) is can be used to make
  for making trade-off decisions; and 
  (H) the history view with automatically saved parameter changes. %   Each parameter changes are automatically saved in the history view (H).
  Hovering the mouse over a historical point reveals %overlays
  its settings and results (in orange).}
    \label{fig:argus_ui}
\end{figure*}

% --------------------------------------------------------------------
\section{Argus User Interface Design} \label{sec:argus}

The \argus interface is organized into: %the 
parameter specification (A--E), simulation output (F--G), and the history view (H) (\autoref{fig:argus_ui}).
%Before using \argus,  users specify 
Users begin by specifying metadata about the independent variables in
%\wm{with} 
a pop-up window % menu %-based interface
(\autoref{sec:dv}).
They can then
% Then, they can 
explore various effect-size scenarios %of effect sizes 
by %specifying 
manipulating the means of the dependent variables for %in
each condition (A).
They can also estimate potential confounds (B); and 
explore how different experiment designs (C--E) influence %influences
the outcome (F--G).
% \alex{@Chat: WM \& Alex were not sure what the following sentence.}
% \alex{The outcome as mean differences (F) and \viewTradeOff (G) can facilitate the decision-making process.}
% \red{These manipulations \wm{support decision-making} %yield output for decision-making: 
% the mean differences (F) and the relationship between power and sample size (G).}
The %This 
history view (H) automatically
%save the progression of exploration, allowing users to 
saves the exploration process and lets users
re-load previous scenarios.
% In the remaining of this section, we will use a running
The rest of this section describes the interface using the example of a 2 $\times$ 2 experiment on how \Medium (\Paper \vs \Screen) and \Layout (\OneColumn \vs \TwoColumn) influences \ReadingTime. 

\begin{figure*}[ht!]
\centering
  \includegraphics{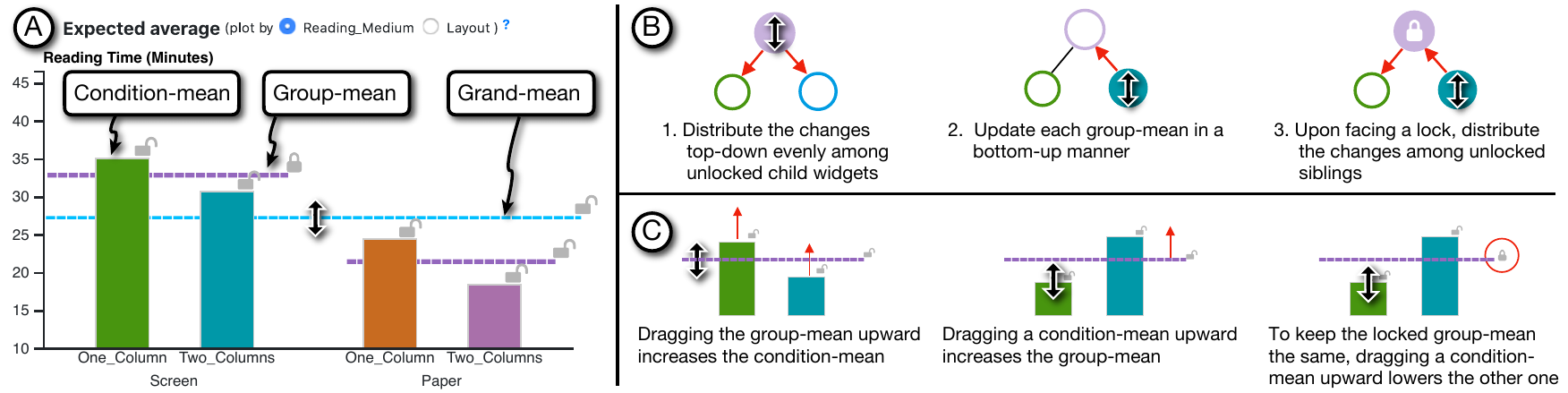}
  \caption{(A) Expected average view: users estimate the mean for each experiment condition;
  (B) Users can lock some means and move others, propagating changes to children, updating group means, or distributing changes to unlocked siblings \changed{(no propagation of changes when both the parent and the sibling are locked)};
  (C) Scenarios show: increasing the condition-mean, increasing the group mean, and locking the group-mean.}
    \label{fig:expectedaverage}
\end{figure*}

%-------------------------------------------------------------------------------
%\subsection{Dependent variable metadata} \label{sec:dv}
\subsection{Metadata} \label{sec:dv}
% Wendy: I know that DV is in the title, but I think it's better to add it the first time it appears in the text as well. Then you can remove it from the title.

To facilitate interpretation of simple effect sizes (C3), 
\argus needs the semantics of the dependent variables.
% These information are typically needed only once 
Researchers supply this information once, 
at the start of the session.
% Because such information is relatively static in each domain, we envision that users should be able to just select the dependent variable, and relevant information are automatically retrieved from a public domain ontology.
Note that, since many domains use a common set of dependent variables, 
such as time and error for VIS and HCI, 
in future, we expect researchers to select relevant dependent variables retrieved automatically from a public domain ontology.
% Such ontology 
Similar ontologies already exist in
%already available for  in the field of
bioinformatics~\cite{Saldatova2006} %.In the field of VIS and HCI, 
, and Papadopoulos et al.~\cite{Papadopoulos2016} have proposed
an ontology that specifies dependent variables for VIS and HCI.
% has already been proposed~\cite{Papadopoulos2016}.
% Therefore, the menu interface we propose below acts only as a temporary substitute.
The current metadata interface is thus a makeshift.

\argus requests %asks for 
the name, unit, expected range, interpretation, and the variability of each dependent variable (DV). 
%We use them to compute initial ranges of the axes in 
\argus computes initial ranges for both axes of
%\wm{and then computes initial ranges for both axes of}
the interactive charts (\autoref{sec:average}), and 
the sliders that adjust
% the range of the sliders to specify 
various confounds (\autoref{sec:confounds}).
\Argus uses the % The 
natural-language interpretation, \eg, "30 minutes is \textit{faster} than 50 minutes", %is used to aid in reading 
to make it easier to read
the pairwise plot (\autoref{sec:pairwise}).

\subsection{\ViewExpectedAverages} \label{sec:average}
\argus uses a direct manipulation interface to determine effect sizes, 
which lets
% We want to enable 
users work with simple effect sizes (T1) and %to be able to 
explore multiple effect-size scenarios. % of effect sizes (C2).
% To address these challenges, we aim to minimize friction in effect size specification by using direct manipulation interface instead of menu interface.
Instead of specifying mean differences, %we let the user 
\argus lets users
specify the expected mean of each experimental condition.
This condition-mean specification lowers user's % lessens
cognitive load %for the users 
because they can flexibly % either 
estimate each condition individually.
%The users  can also ???project???  their estimation and use it as a visual anchor to estimate the difference.
%
% Really unclear. Do you just mean that they can see their estimates in the graphical interface?
%
% \red{Users can also display their estimates for different conditions, 
% which act as visual anchors when estimating differences.} 
% of one condition onto the graphical representation, 

\argus presents the condition-mean relationship as a bar chart (\autoref{fig:argus_ui}.A), and 
\changed{
the bar colors are drawn from the 2D colormap of Bremm et al.~\cite{Bremm2011} by assigning one dimension per variable%
\footnote{We use the Color2D library: \href{http://dominikjaeckle.com/projects/color2d/}{dominikjaeckle.com/projects/color2d/}}%
.
}
%We encode the condition-means with a bar chart, which is a common visualization for sample mean.
% To facilitate visual recognition, the level of each independent variables are encoded in two different visual channels (hue and saturation).
%This condition-mean specification lessens cognitive load for the users because they can flexibly either estimate each condition individually.
%The users can also project their estimation of one condition onto the graphical representation, and use it as a visual anchor to estimate the difference.
Users can %specify their estimation of 
estimate each condition-mean by dragging the bar vertically.
% We use h
Horizontal lines encode the \emph{group-mean} --- calculated 
% which is calculated
from all conditions of an independent variable --- and the \emph{grand-mean} --- calculated from all independent variables (\autoref{fig:expectedaverage}.left).
Despite the potential for within-the-bar bias~\cite{Correll2014}, %the 
encoding the bars keeps condition-mean %bar encoding allows condition-means to be 
visually distinct from the group-means and the grand-mean.
Users can switch the hierarchy level of the condition axis in the bar chart 
%by selecting them in 
via radio buttons.
% This eases visual interpretations according to the Gestalt principle of proximity.  % WEM avoid this sort of general statement.
% \red{To facilitate exploration (C2), we further considered two scenarios when designing this view:}
% \remove{\red{The following two illustrate how our approach facilitates exploration (C2):}}
% \chat{The following are common use cases for expressing effect size.}
We describe two common use cases for expressing effect size:
\vspace{0.5em}

% \noindent
%\textbf{The presence of a main effect} 
\textbf{Main effects} occur when a particular level of an independent variable causes the same change in the dependent variable, regardless of the level of other independent variables.
% is when a level of an independent variable causes the same change in the dependent variable regardless of the level of other independent variables.
For example, a main effect of \Medium on %to the
\ReadingTime
could be that reading on a \Screen is generally slower than reading on \Paper.
To specify this as a main effect, % we
the user would have to drag two bars (\OneColumn and \TwoColumn of the \Screen condition) upward %with the same amount.
by equivalent amounts.
% When an independent variable has many levels, having to drag each bar individually can be tedious.
This becomes tedious when the independent variable has many levels.
\vspace{0.5em}

% \noindent
\textbf{Interaction effects} occur when 
% \textbf{The presence of an interaction effect} is when 
the mean within each group differs according to the level of another independent variable.
Suppose we want to express how the \Layout % column layout
affects \ReadingTime.
As above, we register \Medium as a main effect, but ensure that the group means for \Screen and \Paper remain the same. 

% Continuing the example above, 
% after expressing the main effect of \textsc{Medium}, suppose we want to express how the column layout additionally affects the \textsc{ReadingTime} \emph{while keeping the group mean of the screen and the paper the same} (\autoref{fig:expectedaverage}, left).
If the user changes the $\langle$\OneColumn, \Screen$\rangle$ bar, the group-mean of the \Screen condition will also change.
To keep the same group mean, the user must first remember %have to recall
the group-mean prior, % to the adjustment and move 
and then adjust
the other bars to compensate.

% In these scenarios, users can manipulate 
Both scenarios involve manipulating
multiple conditions simultaneously by dragging group-means and the grand-mean.
% They 
Users can also lock some means while changing the rest, and the system automatically propagates the changes.
However, enabling this interaction technique is tricky because of the hierarchical dependency among these values.

% \red{WEM: I'm having trouble seeing the relevance of this. I'm expecting a straight forward description of how the interface works, with some scenarios to illustrate how Argus handles tricky issues, like accounting for main and interaction effects. The above scenarios are worded as if they are problems, not solutions, and introducing (but not really using) the hydraulic system metaphor at this point seems like a non sequitur.}
% \chat{May be "scenario" is the wrong word. Would "use cases" be more suitable?} Yes, that would work much better. 
% I think the idea for this section is to illustrate the benefits of the Argus design, how it works and how it handles 'extreme' or otherwise difficult cases. A 'scenario' is more like a story, where a fictional users tries out the features of the system to achieve some goal, as a concrete way of explaining the details of the interface.
% \red{We propose a metaphor of a hydraulic system.
% The condition-means, group-means, and grand-means could be seen as water level in a set of interconnected hydraulic cylinders.
% Compressing the water down on one cylinder causes the water to rise up in others.
% If one of the cylinders is locked, the water level rises higher in the rest of unlocked cylinders to compensate.}
% \chat{
% My bad. 
% We can remove the hydraulic metaphor and name that it is a propagation algorithm.
% }

% To me, this is not a 'metaphor' ... Just that we use this algorithm. I get the basic idea of a hydraulic metaphor, but it isn't really covered elsewhere and doesn't help with the explanation.
\argus implements a
% \red{This metaphor} is implemented with the 
propagation algorithm (\aPropagation and 
% The main idea is 
\autoref{fig:expectedaverage}, right).
The relationship between the hierarchy of means is represented as a tree rooted at the grand-mean.
A change to a parent node–––the grand-mean–––is first recursively propagated to the children, \eg group-means and then the condition-mean.
The amount of change is distributed evenly to all unlocked children.
After finishing the change propagation, the update moves %goes 
upward.
If the update reaches a locked parent, the change is distributed to any unlocked siblings.
The propagation algorithm offers users flexibility,
% The Hydraulic Propagation algorithm offers users flexibility,
% in \red{their projection}.
letting them switch seamlessly through different representations at different levels, not only individual conditions, but also main and interaction effects.

% They can seamlessly switch among their representation at the level of the individual condition, at the level of main effect, and at the level of interaction effects.

%  The expected average view (A) allows users to specify their estimation of the mean for each experiment condition. Users can lock some means and move others. The changes will be propagated according to the hydraulic propagation metaphor illustrated in (B), and the full algorithm is in the appendix. Three example scenarios of their behavior are shown in C.

%----------------------------------------------------
\subsection{\ViewPairwise} \label{sec:pairwise}

\begin{figure} %[htb]
\centering
  \includegraphics{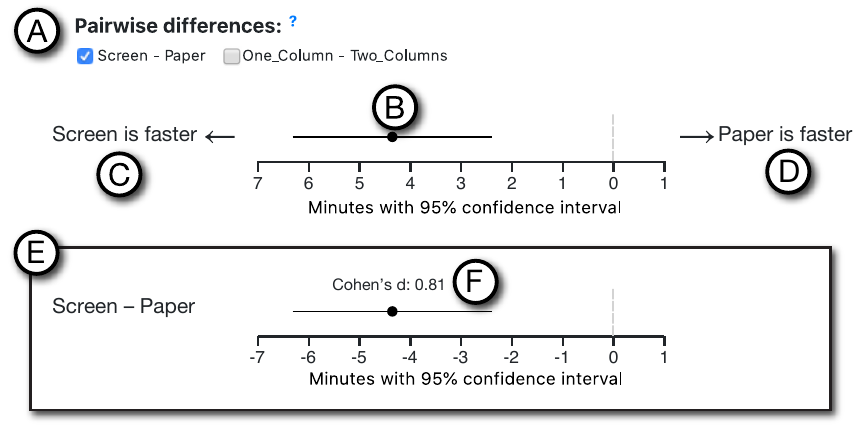}
  \caption{(A) Pairwise-difference view for selecting which effects to include.
  (B) Dancing confidence interval shows the mean differences, with
  (C--D) natural language labels on either side. 
  (E) Holding a Shift key displays labels for mean difference and Cohen's $d$ (F).
  }
    \label{fig:pairwise}
\end{figure}

% The pairwise-difference view lets the user select effects to be included in the simulation (A). 
%   The results are shown as a dancing confidence interval of the mean difference (B).
%   Natural language interpretations are labeled on both sides to aid interpretation (C and D).
%   Users can access an alternative mode (E) by holding a Shift key.
%   In this mode, the label changes to the mean difference, and an additional label showing Cohen's $d$ is shown (F).

To help users evaluate the consequences of their effect size estimates (T2), 
we simulate the data and show the difference between means and their confidence intervals in the \viewPairwise (\autoref{fig:pairwise}).
The horizontal axis shows
% On the horizontal axis is the amount of 
the difference in the original unit of the dependent variable---a simple effect size (C3).
The horizontal axis lists
% On the vertical axis, we list 
all possible comparison pairs. 
An independent variable with $m$ levels can accommodate ${m \choose 2}$ pairwise comparisons. % are possible.
For each pair, we show the mean difference, displayed % encoded
as a black dot, together with its 95\% confidence interval, displayed % encoded
as a black line.
Unlike the bar charts
% We choose this encoding instead of bar charts (as 
used for input (\autoref{sec:average}) % to reduce 
this reduces bias~\cite{Correll2014}.
Although violin plots reduce bias somewhat,
we chose the dot-and-line display because they can fit more lines into a limited space.
% A better debiasing visualization would be violin plots.
% However, the dot-and-line encoding allows more lines to fit in a limited space.
This is crucial when comparing two sets of parameters side-by-side with the history function (\autoref{sec:historyview}). 

%In~\autoref{fig:argus_ui} (F), the topmost error bar shows the difference around 2--6 minutes with a center likely around 4 minutes.
In~\autoref{fig:pairwise}.B, the difference appears to the left
% is on the left side 
of the zero indicator. 
Had we
% If we had 
presented the result on a normal number line, it would have appeared on the negative side, 
and the chart could have been interpreted as: 
% Hence, the chart could have been read as 
``the difference is around minus 4 minutes''. 
Since reading double negatives is
% such double-negative reading could be 
cognitively demanding, 
we present absolute values on both sides of zero on the horizontal axis, 
% Therefore, on the horizontal axis, we present absolute values on both sides of zero 
and add annotations on the left and the right margin (C and D).
This makes it easier for users to interpret, \eg,
% These features could help users to easier reach the interpretation of 
``\Screen is faster for around 4 minutes''.
Users can press-and-hold the shift key to show the normal number line with negative values on the left of the zero, in \autoref{fig:argus_ui}.E.
This mode lets users
% In this mode, we 
change the label on the left margin to present a mathematical difference (``\Screen - \Paper'').
For advanced users, \argus also annotates Cohen's $d$ standardized effect size above each confidence interval.  

In~\autoref{fig:argus_ui}.F, both 
\Screen-\Paper and 
\OneColumn-\TwoColumn are selected.
Suppose we are only interested in comparing reading media because the layouts were included as a nuisance %nuance 
variable.
Deselecting the ``\OneColumn - \TwoColumn'' checkbox might yield a slightly narrower confidence interval for the ``\Screen - \Paper'' difference.
The reason for this improvement is that the difference between the two layouts is slightly smaller in the \Paper condition (\autoref{fig:argus_ui}.A),
\ie there is an interaction effect.
%\wm{\ie an interaction effect.}
% In other words, there is an interaction effect.

Since \argus shows simulated data instead of real data collected from an experiment, we 
need to ensure that users are aware of the uncertainty generated by  %would like to make users aware of the uncertainty from 
the simulation.
We thus % Alex: @Wendy, is it a stylisitic choice to write "We thus..." instead of "Thus, we..."?
% Therefore, we 
use the \emph{dance of the CIs}, a time-multiplexing approach that shows the results of multiple simulations in the same figure~\cite{cumming2012, Dragicevic2019}.
The animation runs in 2 fps, to allow the user to notice changes between frames~\cite{trick1994}.
An alternative to the dance animation is a forest plot 
that displays all confidence intervals from the simulation
% in which all confidence intervals from the simulation appear 
next to each other, 
with a diamond shape to summarize them~\cite[Chapter 9]{Cumming2017}.
% and a summary of these intervals is shown in a diamond shape

We chose the dance because it uses less screen space, and motion is a strong visual cue.
Even when the user focuses somewhere else on the screen, the animation is registered in their peripheral vision. 
\changed{In addition, users can pause the animation and navigate individual frames by the left and right arrow keys on the keyboard.}

\begin{figure}
\centering
  \includegraphics{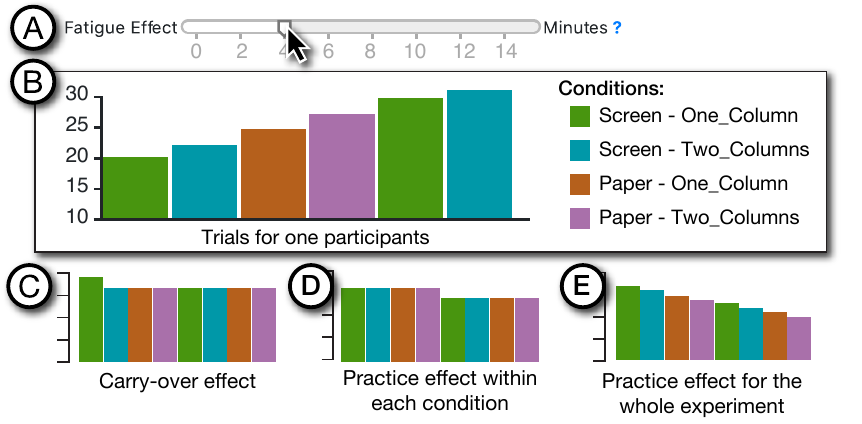}
  \caption{(A) Adjusting the `fatigue' confound effect level
  (B) displays its corresponding influence on the data, 
  as well as 
  (C) carry-over effects, (D) practice effects per condition and (E) for the whole experiment. 
  %\remove{Note: For the purposes of illustration, (D) uses a different experiment design.}
  }
    \label{fig:confounds}
\end{figure}

% Moving the slider for a confounding parameter such as the fatigue effect (A) shows a pop-up (B) to visualize how the effect will systematically influence the data.
%   Such visualization is also shown for three other confounding parameters (C--E).
%   (For the practice effect within each condition, the experiment design is different from others for the purpose of illustration.)
  
%=======================================
\subsection{Exploring Trade-offs}
At each effect-size scenario, users can increase power by adding more participants, increase the number of trial replications in the counterbalancing design, %collecting more replications of data
or both.
Some experiments may be constrained by participant fatigue and need to limit the duration, 
whereas for
% but in 
other experiments, the cost of recruiting additional participants may outweigh the drawbacks from the fatigue effect.
Argus lets users explore how different experiment design scenarios
% We should enable users to explore how scenarios in experiment design 
and confounds can influence power (T4), 
as shown % These issues are addressed by three parts of \argus user interface shown 
in~\autoref{fig:argus_ui}.
Users estimate levels for %the amount of 
each potential confounding effect (B) and select an experiment design parameter accordingly (C--E).
They explore how the trade offs change based on
% Then, they observe the change in the trade-off between
sample size and power (G), 
and can revisit and compare earlier explorations
% Past explorations can be revisited and compared 
with the \viewHistory (H).

%----------------------------------------------------
\subsubsection{Confound Sliders} \label{sec:confounds}

Confounding effects can be specified by sliders (\autoref{fig:argus_ui}.B).
When users drag a confound slider, \argus shows a pop-up overlay to preview its effect (\autoref{fig:confounds}).
The pop-up is a bar chart showing how the measurement of the dependent variable (vertical axis) could change along with the experiment trials (horizontal axis).
The order of trials and the effects are calculated based on the choices in the \viewExperimentDesign (\autoref{sec:exp_design}).

Four types of confounds are of interest in power analysis \changed{\cite{Lazar2017Chap3}} .
For readability, we will explain each of them in terms of reading time.
Increasing the \emph{fatigue effect} (\autoref{fig:confounds}.A) would cumulatively increase the reading time for each subsequent trial (\autoref{fig:confounds}.B).
The \emph{carry-over effect} (\autoref{fig:confounds}.C) occurs when the user is unfamiliar with the task itself:
Their performance is worst in the first trial, but 
gradually improves over 
% once they have done the task once, the performance becomes better in all of the 
subsequent trials, regardless of the experimental condition.
The practice effect has two variations:
The \emph{within-condition practice effect} (\autoref{fig:confounds}.D) represents improvements resulting from the participants' familiarity with each experimental condition. Thus, improvement in one condition does not influence subsequent trials in other conditions.
The \textit{whole-experiment practice effect} (\autoref{fig:confounds}.E) results from users' familiarity with the task, regardless of experimental condition.
This is the opposite of the fatigue effect.
A participant in our think-aloud study (\aThinkAloud)
pointed out the difference between these two practice effects, and we plan to incorporate the whole-experiment practice effect in the next version of \argus.

The confound pop-ups use a bar chart to encode the level of the dependent variable.
We take advantage of the Gestalt law of similarity to let the user associate the color-coding of conditions to those in the \viewExpectedAverages.  
% WEM I'm not found of throwing in things like the Gestalt law for this ... sounds like you're trying to impress me, the reader, but are not really sure of the underlying Psychology.
% \alex{@CHAT: there is another comment in the latex.}
%\red{An alternative visualization is a stacked bar chart that incorporates all effects into a single chart.
% We did not use the stacked bar chart because when the bars are unaligned, it is difficult to spot the pattern that each individual confound contributes.}
Future versions of \argus could include a more advanced interaction technique that lets users specify a range or a probability distribution for each confounding variable.

\argus uses the dependent variable metadata (\autoref{sec:dv}) to determine the range for each slider.
The direction of the available values depends upon which direction users specify as the ``better'' direction.
For example, in~\autoref{fig:argus_ui}.B, the variability is set to $\pm5$ minutes, and the interpretation is specified as ``slower is better''.
These settings create a fatigue-effect slider ranging from 0--15, and a practice-effect slider ranging from -15--0.
All sliders are initially set to zero to represent no confounding effects.
\argus also provides an additional slider for specifying variations across participants.

%----------------------------------------------------

\subsubsection{\ViewExperimentDesign} \label{sec:exp_design}

The effect of confounds such as the fatigue effect could even out across participants if the experiment is properly counterbalanced.
In the running example, the experiment has four conditions.
A complete counterbalancing would require covering the $4! = 24$ possible orderings of the conditions, 
which would in turn require 
% and therefore, 
recruiting a \emph{multiple} of 24 participants. 
Alternatively, users might consider using a standard Latin Square design, 
which addresses % WEM This is a bit less strong.
% rules out  
the order effect between adjacent trials.
This Latin Square design requires only multiples of four participants, allowing for greater flexibility in the sample size.

Recruiting fewer participants than required multiple may lead to an imbalanced experiment, 
and affect both the observed effect and power.
Finally, users could collect several replications of data from each participant.
This number of replications influences the trial table, and thus influences how the confounding effects contribute to the data.

In the field of HCI, several tools exist for counterbalancing design~\cite{Eiselmayer2019, Mackay2007, Meng2017}.
Eiselmayer et al.~\cite{Eiselmayer2019}'s interview study suggests that counterbalancing design and power analysis are performed in two separate loops.
We envision that users should use one of these tools to come up with experiment design candidates.
Then, these candidates can be imported to \argus.
For these reasons, we present a minimal user interface for counterbalancing design: a drop down list for selecting the counterbalancing strategy (\autoref{fig:argus_ui}.C) and two sliders for the number of replications (D) and the number of participants (E). 
These controls work together with the \viewTradeOff and \viewHistory.

%----------------------------------------------------
\subsubsection{\ViewTradeOff} \label{sec:tradeoff_view}

The \viewTradeOff (\autoref{fig:argus_ui}.G) is the heart of power exploration (T3). 
It visualizes the outcome of the adjustments in \viewExpectedAverages, \viewConfound, and \viewExperimentDesign.
The visual encoding is based on the chart relating power vs. sample size, commonly used in statistics textbooks, \eg~\cite{Scheffe1959}.
The sample size appears on the horizontal axis and the power on the vertical axis.
The current selection of the sample size is represented as a dot, 
and the relationship between these two parameters are
displayed as a black curve.
We used this encoding despite the fact that the underlying data is discrete---the sample sizes are integer---because curves facilitate interpretation of the local rate of change~\cite{Cleveland1986}, which is usually the case when researchers %decide on the
assess power trade-offs.

Touchstone2~\cite{Eiselmayer2019} enhanced this textbook chart by automatically showing the confidence band around the current parameter set, which
% The confidence band 
was calculated from a single ``margin'' parameter.
In \argus, variations in power can originate from any of a combination of multiple sources, \eg, effect size or confounds, 
making it difficult to determine which are associated with the confidence band.
% It would be unclear which of the sources of the variation that the confidence band is associated with.
% One could use the power calculated from individual datasets from the simulation to form the confidence band. (BUT WHY NOT??
% Touchstone2 also only shows the sample sizes that are valid for the selected counterbalancing design (\eg, a multiple of 4 participants in the example above) as dots on the line and the ticks on the horizontal axis. 
% We exclude this visualization because of WHAT??

\argus enhances this chart in two ways:
First, Users can switch the horizontal axis between the sample size and the number of replications. 
Setting the axis to the sample size shows the number of replications annotated on the right end of the power curve.
This switch could be used when the sample size faces a stricter constraint than the number of replications, or vice versa.
In~\autoref{fig:argus_ui}(G), suppose the resource constraint allows the recruitment of a maximum of 24 participants, which results in the power of 0.7.
Users can now consider the trade-off between the number of replications and power.
% NOTE: In hindsight, it would be better to show two charts next to each other for the two pairs.

Second, \argus shows the chart individually for each of the pairs of independent variable levels, \eg,~\autoref{fig:argus_ui}.G, shows "\Screen - \Paper").
Users can change the pair with a drop-down menu.
\argus shows a warning if any pairs produce lower power than the current pair.
The user can also select the "Minimum power" option to always display the pair with the lowest power.
Although this pair-selection is also present in the \viewPairwise, the selection in \viewTradeOff is independent:
Switching it does not trigger a simulation.
This independence allows the user to explore nuisance factors without changing how the confidence interval of differences is calculated.

%----------------------------------------------------
\subsubsection{\ViewHistory} \label{sec:historyview}

The \viewHistory (\autoref{fig:argus_ui}.H) ties together all above-mentioned views to enable exploration of scenarios in light of uncertainty from effect size estimation and confounds (T4).
\Argus thus improves on other power analysis systems that force users to record each scenario's output before manually comparing them.
% It improves on most of the power analysis systems \wm{that force} users to record the output of each scenario \wm{if they } manual comparison 
(\autoref{sec:related_work}).

Each step of parameter adjustment is  recorded automatically in an abstract tree.
The root of the tree is the initial setting of zero effect size with no confounding variables.
The tree is visualized on a two-dimensional cartesian coordinate with the vertical axis showing the power.
The horizontal axis shows the depth of the node from the root.
Each node is encoded as a white circle with black outline, and it is connected to its parent node with a line.
The current node is encoded in a black circle to associate it to the the dot in the \viewTradeOff with the Gestalt principle of similarity.
Adjusting a widgets in the views mentioned above creates a child node.
Clicking on a past node restores its parameters all other views.
The restoration excludes the selections in the \viewTradeOff to enable users to retain their current focus, as described in~\autoref{sec:tradeoff_view}.
During exploration, it is likely that only a few nodes will be of interest. 
Users can mark/unmark a node by clicking a button. An additional concentric outline circle is added to each of the marked nodes. 

In addition to restoring the parameters, users may hover their mouse cursor over a node to preview its parameters and output.
The preview values are shown in orange, simultaneously with the values of the current node in black (\autoref{fig:argus_ui}).
We use juxtaposition and superposition faceting techniques.
These two techniques were analyzed in Javed et al.'s survey of composite visualization~\cite{Javed2012}.
Their analysis found that for tasks that focus on direct comparison in the same visual space, superposition is more effective than juxtaposition.
For the \viewTradeOff, since decisions about sample size usually take place around the few crucial values (see C2 and ~\autoref{fig:power_decision}), we superpose the curves.
For the \viewConfound and \viewExperimentDesign, the sliders and the drop-down list, preview values are also superposed.
For the \viewExpectedAverages, however, both superposition and juxtaposition would be appropriate.
Here, superposition allows the bars representing the current state to provide a stable visual anchor.

For the \viewPairwise, the uncertainty communicated by the animation would be muddled when two superposed confidence intervals overlap. 
Therefore, we juxtapose the preview error bars side-by-side (\autoref{fig:argus_ui}.F).
For the \viewHistory itself, we highlight nodes and edges in the current branch during preview.

We also decided to limit the comparison to two nodes---the current node and the preview node---to reduce visual complexity.
A pairwise comparison of historical nodes together with the marking functions allows users to gradually narrow down the parameter choices.

\subsection{\changed{Scaling the Design for More Complex Experiments}}
Our prototype supports within-participants designs with two independent variables.
More complex experiment designs may have more than two independent variables, and each independent variable could have more levels. 
Only two views will be affected:
The \viewExpectedAverages could present more levels by incorporating the fish-eye technique~\cite{Rao1994}.
To address more independent variables, the system should allow the users to reorder the hierarchy in the horizontal axis---\eg, by drag-and-drop.
Users should also be able to exclude some of the independent variables from the axis, which will summarize several bars of the same level into one, which further reduces the visual complexity.
As for the \viewPairwise, scrolling and panning could be necessary to handle the increased number of pairs.
When their effect sizes are very different in the magnitude or sign, the comparison could be broken down into subsets, presented in separate windows.

%% file: _sections/06-1_implementation_details.tex
%!TEX root=../_main.tex

% --------------------------------------------------------------------
\section{Implementation Details} \label{sec:implementation}
% --------------------------------------------------------------------
\begin{figure*}[ht!]
\centering
  \includegraphics{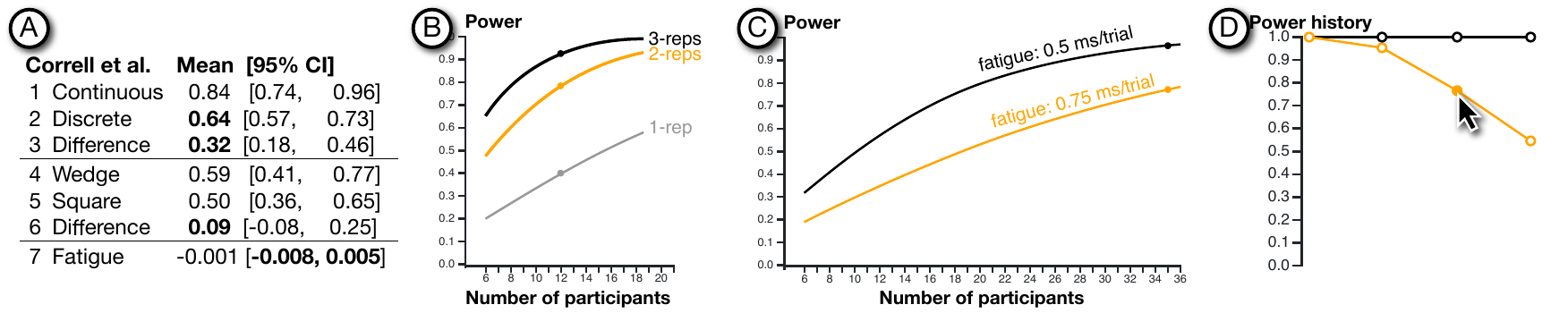}
  \caption{
  (A) Releveant error estimates based on Correll et al.'s data;
  (B) The power is plotted against the number of participants 1-, 2-, and 3-replication scenarios. 
  (In Argus UI, only the maximum of two curves are shown at a time during interactive comparison.)
  (C) Power trade-off curve of three-replication with the fatigue effect of 5 ms (in black) and 7.5 ms (in orange). 
  (D) The History view showing two branches: three-replication (in orange) and two-replication (in black).}
  \label{fig:usecase1}
\end{figure*}
\changed{\Argus was written} in HTML and JavaScript. 
We used D3.js\footnote{\href{https://d3js.org}{d3js.org}} for interactive visualizations.
Experiment designs are implemented in the TSL language and trial tables are generated on the client-side with the TSL  compiler~\cite{Eiselmayer2019}.
Statistical calculations are implemented in R\footnote{\href{https://r-project.org}{r-project.org}}, and Shiny\footnote{\href{https://shiny.rstudio.com}{shiny.rstudio.com}}.
We used a MacBook Pro (2.5GHz, 16GB memory, MacOS 10.14) for all benchmark response times.

\changed{
To enable interactive exploration in \Argus, we make the following three implementation details that differs from standard statistical procedure for \textit{a priori} power analysis and post-study statistical analysis.
}

\subsection{Monte Carlo Data Simulation}  \label{sec:simulation}
Power can be calculated from an $\alpha$ probability value, a standardized effect size, and a sample size.
However, incorporating confounds, \eg, a fatigue effect, is % difficult
analytically complex (C4). % Alex: @Wendy, if case you see this. Is it a stylistic choice to put complex before analytically instead of the other way around?
%WM: Actually, you can do it either way, probably because of the verb 'to be". If it were another action verb, I would try to avoid putting an 'ly' adverb before. Interesting subtleties here. I reread it and it sounds better this way.
Instead, we use a Monte Carlo simulation, based on algorithm 1 of~\cite{Zhang2014}:
First, a population model is created programmatically, based on an estimate of the mean and the standard deviation (SD) of each condition.
From this population, we sample data sets and use them 
to calculate statistics.
%for statistical calculation (see section 4.2) as described in the following section.
The Monte Carlo paradigm has been shown to be robust for tricky cases such as data that are not normally distributed, missing data, or
 % mixture of 
mixed distributions~\cite{Muthen2002,Schoemann2014,Zhang2014}.

We extend the algorithm to incorporate confounding variables: 
First, we obtain a trial table for the specified experiment design from the TSL compiler.
Based on the trial table's structure, we generate each 
%of the confounding effects as users specified in the UI 
confounding effect specified by the user in the interface (\autoref{sec:confounds}).
For example, a two-second fatigue effect % of 2 seconds on the
for movement time 
cumulatively lengthens each subsequent trial by two seconds.
% makes each of the subsequent trial cumulatively longer by 2 seconds.
%Other confounding effects are generated in a similar manner. WEM this is redundant
All confounding effects are added to each simulated data set before 
calculating statistics.
% statistical calculation. 
Data simulation and confounding calculations are vectorized.
On average, 
we can generate
% generating 
a data set with 50 participants and 
10 replications with all confounding effects in place,
in less than 30 ms on our benchmark machine.

\subsection{Making Power Calculation Responsive}
Calculating statistical power is computationally expensive because it requires a numerical integration between % Alex would say "between"
two overlapping probability distributions (see Fig.~11 of~\cite{Eiselmayer2019}).
Furthermore, post-hoc power calculation uses an \emph{observed effects size} from the data, which may differ from the input effect size due to confounding effects.
To calculate observed effect sizes, we must fit a general linear model for each data set.
In normal statistical analysis, such model-fitting is done only once, so results appear almost instantaneously.
However,
plotting the chart of sample size and power
% \wm{plotting the sample size and power chart} 
% to plot the chart of sample size and power
(\autoref{fig:power_decision})
requires one calculation per simulated data set.
%these calculations are required once per simulated data set.
By default, \argus generates 1000 data sets for each sample size. 
Here, we show the sample size from 6 to 50.
On our benchmark machine, the entire calculation takes around two--three minutes.

To ensure the responsiveness of the user interface,
% \wm{To make the user interface fully responsive,} % WEM check slightly different meaning here
% To allow the user interface to be responsive, 
we first % give the
approximate the observed effect size with a pairwise Cohen's $d$ calculated with the \texttt{pwr.t.test} function from the \texttt{pwr} package~\cite{Champely2018}.
The average turn-around time is 200 ms.
Model-fitting results are sent progressively to 
% The results based on model-fitting are progressively sent
the user interface, 
which updates accordingly. % WEM again, slightly different but I think better
% which is gradually updated.
\changed{
We further ensure responsiveness, we also make further tweaks in the communication between R, Shiny, and Javascript as detailed in \aArchitecture.
}

\subsection{Statistical Model and Pairwise Difference Calculation} \label{sec:pairwise_model}
After modeling participants as a random intercept, we
derive the observed effect size and the pairwise difference in terms of means and confidence intervals from mixed-effect models.
% , with participants modeled as a random intercept.
(See Fry et al.'s~\cite{Fry2016} HCI statistics textbook for more details on the model choice.)
% Details on the model choice are discussed in a HCI statistics textbook~\cite{Fry2016}.
\argus automatically formulates a mixed-effect model and a contrast matrix for generalized linear hypothesis testing, based on the user's choice of the condition pairs of interest (\autoref{sec:pairwise}),
% Alex flipped the sentence so it is less front heavy.
% Based on the user's choice of the condition pairs of interest (\autoref{sec:pairwise}), \argus automatically formulates a mixed-effect model and a contrast matrix for generalized linear hypothesis testing.
We use the \texttt{lme4} package~\cite{Bates2015} for model fitting and the \texttt{multcomp} R package~\cite{Hothorn2008} for the test.
Confidence intervals are calculated with a single-step adjustment with the family-wise error rate set at $\alpha = .05$.

%% file: _sections/06-2_use_case.tex
%!TEX root=../_main.tex

% --------------------------------------------------------------------
\section{\changed{Use Case}} \label{sec:usecase}
% --------------------------------------------------------------------

To demonstrate how to use Argus, we draw an example from a study on color ramps from Smart et al.~\cite{Smart2020}---of which the study plan could have been informed by a similar study by Correll et al. ~\cite{Correll2018}. 
Additionally, both studies made their data publicly available, allowing us to derive additional information for planning and testing. 
We first describe the background of both studies—which constrains the parameter space to be later explored with Argus. 
To aid cross-referencing, we highlight relevant values in \textbf{bold}. 
Calculation details are provided with R code in supplementary S2.

\subsection{Background}

Smart et al. propose to generate color ramps based on a corpus of expert-designed ramps by using Bayesian-curve clustering and k-means clustering. 
Their experiment compared four types of ramps (\cBayesian, \cKMeans, \cDesigner, and the baseline \cLinear) in three visualization types (scatterplots, heatmaps, choropleth maps), in a total of 12 conditions. 
In each experimental trial, study participants are asked to identify a mark on the visualization that matches a given numerical value. 
They measured errors and aesthetic ratings. 
Because a comparable aesthetic data were unavailable in prior works, this use case focus only on the errors, which is defined as $|v_\text{given} - v_\text{selected}|$.

\changed{%
To plan their study, Smart et al.'s study could have leverage information from Correll et al.'s experiment%
\footnote{Although Smart et al. mentioned that their study was similar to ~\cite{Gramazio2017}, the latter concerns categorical palettes rather than quantitative color maps.}%
.
The latter used the same identification task, albeit only heatmaps are used as the visualization.} 
Their study investigated how color ramps can be used to encode both values and uncertainty. 
Although their experiments have different conditions compared to Smart et al.'s, two of their results are relevant: (1) the significant difference between continuous vs. 
discrete color map, and (2) the absence of a statistically significant difference between wedge-shaped vs. square-shaped color legend. 
The former can be used as an upper-bound and the latter as a lower-bound for the effect sizes. 
Since Correll et al.'s accuracy was defined differently from Smart et al.'s error, we use Correll et al.'s data to calculate the errors—which result in the statistics shown in~\autoref{fig:usecase1}.A.

In addition to the effect sizes, we also retrieved the duration information. 
In each trial of the relevant experimental condition, participants took 8.5 seconds. 
Since the stimuli of Smart et al.'s study was four times larger, we extrapolate \textbf{each trial to take 34 seconds}. 
In Correll et al.'s study, the median session duration was 13.5 minutes. 
We also analyzed the data for the fatigue effect and found it negligible with the estimate in ~\autoref{fig:usecase1}.A, row 7.

Smart et al. recruited \textbf{35 expert designers} as their study participants; we use this number as a maximum number of participants. 
On the opposite, we consider \textbf{12 as a minimum number of participants} based on a rule of thumb~\cite{Eiselmayer2019}. 
Since the participants were experts, they might be less willing to participate in a long study. 
Therefore, we constrained the longest session duration to 30 minutes. 
Leaving 5 minutes aside for instruction and informed consent, this results in \textbf{the maximum of 3 replications} ((25 minutes $\times$ 60 seconds) $\div$ (12 conditions $\times$ 34 seconds) = 3.6, rounding down) We used the randomized counterbalancing according to Correll et al.'s design. 
We will aim for power above 0.8—according to Cohen's recommendation ~\cite[p. 56]{Cohen1988}.

\subsection{\textit{A priori} Power Analysis}
\changed{%
In the following scenario, the goal of the researcher%
\footnote{The researcher will be further referred to as a gender-neutral ``he''.}%
 is to determine the sample size (number of replications and number of participants) for his experiment.
As mentioned above, these decisions are constrained by the total duration of the session, maximum number of participants, and potential for confounding effects.
}
The exploration starts with the upper-bound and lower-bound scenarios and proceeds to explore a potential fatigue effect.

\subsubsection{Upper-bound Scenario}
He started with 12 participants and 1 replication. 
He moves the grand mean to 0.64 and the group-means of conditions other than the \cLinear to 0.32 \textbf{(T1)}. 
These values are from Correll et al. discrete conditions (\autoref{fig:usecase1}.A, row 2), and its difference to the continuous conditions (\autoref{fig:usecase1}.A, row 3). 
On the Power Trade-off view, the researcher sees that the power of the effect between \cLinear – \cDesigner pair almost 1.0, which is very high---indicating that if the effect size is large, only 12 participants would be adequate \textbf{(T3)}.

\subsubsection{Lower-bound Scenarios}
He moved the group-mean of the \cDesigner condition to 0.55 (from ~\autoref{fig:usecase1}.A, row 6). 
The power drops to around 0.4. 
One way to address this is to increase the number of replications to 2 and 3, resulting in the power of 0.7 and 0.9 respectively \textbf{(T3)}. 
He hovers his mouse cursor on the history nodes to superpose the power curves in Power Trade-off trends (\autoref{fig:usecase1}.B). 
According to the curve, for one- and two-replication designs, adding participants would dramatically increase power. 
However, for 3-replication setting already have relatively high power \textbf{(T3)}.

\begin{figure}
\centering
  \includegraphics{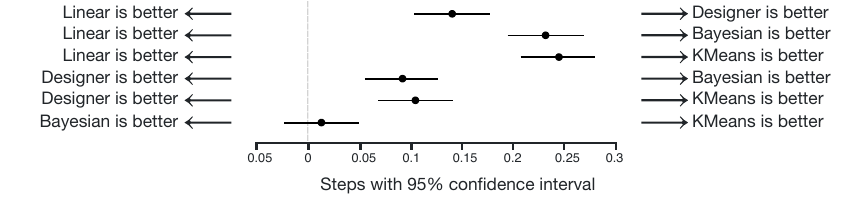}
  \caption{%
  The pairwise difference plot from the case study.
  }
  \label{fig:usecase2}
\end{figure}

Naturally, the researcher would hope that the \cBayesian and \cKMeans will be better than \cDesigner ones. 
However, he does not know \textit{a priori} which of the two algorithmically-generated ramps will be better. 
To reflect these beliefs, he moved both \cBayesian and \cKMeans to 0.46 \textbf{(T1)}. 
These values reflect a small effect when comparing with \cDesigner condition. 
However, when comparing with \cLinear condition, the difference is sizable. 
In the Power Trade-off view, he switches to the pair Designer – Bayesian and found the power to be above 0.8 \textbf{(T3)}. 
The pair-wise difference (\autoref{fig:usecase2}) shows the difference between all pairs except \cBayesian vs. \cKMeans to be larger than zero. 
Also, the difference between \cLinear and the two algorithmic conditions is larger than between \cLinear and \cDesigner. 
Results like these matches the researcher's expectation; therefore, he marked this point in the History view as a plausible design \textbf{(T2)}.

\subsubsection{Fatigue Effect Scenarios} 
From the scenario above, the total duration of a study session is 20.4 minutes (3 replications $\times$ 12 conditions $\times$ 34 seconds/trial). 
This duration is longer than Correll et al.'s median of 13.5 minutes. 
Therefore, it is possible that the fatigue effect \changed{may have influenced} the experiment. 
To explore its impact, he adjusts the fatigue effect to 5, 7.5, and 10 ms per trial---according to ~\autoref{fig:usecase1}.A, row 7---and found that the power drops very low \textbf{(T4)}. 
Therefore, he changes his exploration strategy to determine how much of the fatigue effect could his study design tolerate at the maximum number of participants of 35. 

He set up the 35 participants without any fatigue effect as a starting point and mark it in the History view. 
Then, he creates two branches of scenarios: two- and three-replications. 
In each branch, the explore the three levels of fatigue effects mentioned above \textbf{(T4)}, resulting in ~\autoref{fig:usecase1}.D. 
The two-replication scenarios seem not to change the power much  \textbf{(T3)}---and hence robust to the fatigue effect. 
However, collecting two data points per condition could be susceptible to outliers. 

On the other hand, in the three-replication branch, the power reduces dramatically as the fatigue effect increases \textbf{(T3)}. 
By selecting one node (fatigue: 5 ms/trial) and hovering on another (fatigue: 7.5 ms/trial), he can compare the two corresponding curves in the Power Trade-off view (\autoref{fig:usecase1}.C). 
From the orange line in this chart, he can see that if the fatigue effect is higher than 7.5 ms, the experiment will need more than 35 participants to achieve power at least 0.8. 
He could not effort this scenario \textbf{(T3)}.

To decide between the susceptibility to outliers or the fatigue effect, he could run a pilot study to assess the impact of the fatigue effect with the three-replication setting. 
If the fatigue effect is 0.5 ms/trial or lower, an experiment with only 22 participants would be adequately powerful. 
We validated this potential choice by a simulation that resamples data from Smart et al.'s result and found that recruiting only 22 participants are likely to generate similar outcome as those reported in Smart et al.'s paper.
The simulation details is provided in supplementary S2.

%% file: _sections/07_think-aloud_study.tex
%!TEX root=../_main.tex

% --------------------------------------------------------------------
\section{\changed{Think-aloud Study}} \label{sec:thinkaloud}
% --------------------------------------------------------------------

To better understand how \argus users could be used in power analysis, we conducted a formative study that aims to answer the following research question:
What insights can researchers gain from being able to interactively explore the impact of design choices for their experiments.
The study was preregistered (\href{https://osf.io/2nh4v/?view_only=2207553a2ec94eaa8eeba6f3b9d11e63}{Anonymized URL}) and is fully described in \aThinkAloud.
This section provides a summary.

\subsection{Method Summary}

\textbf{Participants:} 
Nine researchers in HCI and/or VIS participated in our study.
Five of them were experienced researchers who has conducted three or more experiments. 
They were either senior scientists (post-doc or higher), and one was a senior-year Ph.D. student.
The rest of them were Ph.D. students or post-docs who had learned about experimental method, but had planned less than three experiments.
Henceforth, the participants in our study will be referred to as ``users'' To avoid confusion with the ``number of participants'' term in Argus.

\noindent
\textbf{Task and procedure:} 
We used a think-aloud protocol where users voice their observations and reasoning~\cite{Lewis1982}.
The users watched a video explaining \argus and relevant concepts in experiment design and statistics.
Then, they used \argus to determine a sample size for a Fitts's law experiment based on a summary of prior findings.
At the end of the session, we interviewed and asked them to rate their experience.

\noindent
\textbf{Data analysis:} 
We recorded users' screen and audio think-aloud and interview responses.
We performed a qualitative analysis with bottom-up affinity diagramming with the focus on insights~\cite{Saraiya2005}.

\subsection{Selected Results}
\changed{%
Overall, the majority of the users reported that they have gained new insights about experiment design:``\textit{the preview is very useful to understand the confound effects.}''(\pNine).
\pSeven, \pEight were not familiar with carry-over effect and practice effect but they expressed their understanding of the difference between these effects when they saw the previews. 
Five users applied their experience in conducting experiment to consider potential confounds. 
%P2E P3N P4E P6N P8N
For example, \pEight said \textit{``adding more replications can yield higher power but participants may be tired [so] I need to increase the fatigue.''} after increased the number of replications.
}

The influences of the number of replications and participants to power were explicitly observed: 
\textit{``The power is very high now. I am going to tweak replications and participants to see how power is going to change [...] reduce the number of participants, power drops down. It makes sense''} (P4).
Participants also interpret the characteristics of the curve in \viewTradeOff: 
\textit{``The power get stabled after a certain number of participants. The current number of participant is a bit too much. We can reduce the number''} (P5).

However, three of the expert users were initially puzzled why changing the practice effect slider did not influence the mean-differences nor the power.
The study moderator had to point out that the effect was prevented by the Latin-square counterbalancing, or because only one replication was used.
This result suggests an opportunity to improve users' awareness when causal links are muted by a moderating parameter. (See the transition matrix in Appendix D for how users inferred the causality between power analysis parameters)

Five users tweaked expected confounds and observe how the power of adjacent nodes in the \viewHistory gradually changes. 
Four users repeatedly used the hover function to preview the difference.
Two expert users use the branching to explore multiple strands of parameter configurations.
These behaviors show that the \viewHistory successfully facilitates the exploration of statistical power.

%% file: _sections/09_lesson_learned.tex
%!TEX root=../_main.tex

% --------------------------------------------------------------------
\section{Lessons Learned} 
% --------------------------------------------------------------------

We have went through many cycles of design, prototyping, and testing.
It was fascinating to see how the context of use (statistics) influence users' expectation and behavior when interacting with \argus.
We would like to share three lessons:

\textbf{L1: Enabling visual exploration and close-loop feedback generates curiosity about causal relationships.}
The \viewHistory enables users to compare different scenarios.
Our task analysis shows that the focus of comparison is the relationship between the statistical power and sample sizes.
Therefore, in an early version, hovering the mouse cursor on a historical node showed the differences only in the \viewTradeOff and the \viewPairwise.
For other views, the input parameters were temporarily reverted back to the  state of the historical node.
For example, the knob of confound sliders is positioned at the state of the historical node.
However, users who tested this version of \argus are curious to see the differences in the input parameters as well.
We surmised that the immediate feedback from simulated data and the  the affordance for parameter exploration piqued their curiosity of the causal relationship between each of the input parameter to the power.
This evolution of users' need is another evidence that visualization design is essentially iterative.

% In the initial design of \argus, users were able to compare different scenarios based on their statistical power with the \viewHistory.
% However, the hovering over previous versions reverted the state of all input controls back to that version.
% With this design, we tried to improve upon related tools for power analysis.
% Comparing power at multiple effect size scenarios (C2) was one of the challenges we addressed.
% In a subsequent design iteration, we changed the \viewHistory interaction so that users can compare whole effect size scenarios \textit{including} their parameter configuration.
% By using superimposing layers, users are not only able to explore but to understand the complex relationship between the input parameters, and the statistical result and power.
% Newman and DeCaro~\cite{Newman2019} found that exploration can be beneficial in comparison to direct instructions for the learning success.
% Our participants seemed eager to learn about this relationship, especially when they were not able to make sense for their manipulation.
% For example, the practice effect confound only affects experiments where a condition is repeated consecutively.
% This contradicted the mental model after which participants increase their performance over the course of the experiment, \ie a reverse fatigue effect.
% While the practice effect slider was by far the most explored input control (\autoref{fig:interaction_average}), none of the participants were able to make that inference (\autoref{fig:table_interactions}).

% Alex end

%--------------------------------
\textbf{L2: The ease of verbalization could be important for integrating the domain knowledge to interpret visualized data.} 
In \viewPairwise, we used points and error bars to visualize the results of simulation.
An early version of \argus shows output in terms of arithmetical difference (\autoref{fig:pairwise}, E).
Some users struggled to understand the effect when the difference falls on the left of the zero.
To address this problem, we changed the default display mode to show natural language labels (\autoref{sec:pairwise}).
After this addition, we did not observe this difficulty.
Automatically-generated verbal description of visualization has been shown to help users understanding statistical test procedures~\cite{Wacharamanotham2015} and to support understanding of machine-learning models~\cite{Hohman2019}. 
We conjecture that, for the tasks that requires users to combine visual interpretation with their domain knowledge, verbalization is important for the users to successfully integrate visual processing with their knowledge. 

%--------------------------------
\textbf{L3: When asking for a ballpark, avoid precise terms.}
\argus needs a rough approximation of the standard deviation (SD) of the population of the dependent variable to initialize the range of the confound sliders.
This initial value is important to set an appropriate range and granularity of the sliders.
However, it does not need to be precise.
After the sliders are initialized, users can come back to change this value any time to expand or contract the range of the slider.
In an earlier version, the UI simply asked the user to input a number into a text field with the label \textit{"Approximated SD"}.
This question turned out to be difficult for people we pilot-tested the software with.
Some of our colleagues even invested time to lookup research papers in order to give an accurate value.
In a later version of \argus, we reworded it to \textit{"Variability"}, which is a broader term that could be understood as, \eg, SD, variance, or simply a range.
This change seems to lower the users' anxiety and proceed to use \argus faster.
We conjecture that the context might have also putting the users unnecessarily on guard.
Pilot testing with users are helpful to identify such unintended barriers, especially for the choke points of the task flow.

%% file: _sections/08_Discussion.tex
\section{Discussion} 
% --------------------------------------------------------------------

%\textbf{\textit{Notes to add to the discussion section}}
% There is an on-going debate about the fixed $\alpha$ = .05 convention [Benjamin2018 and Amrhein2019].

\argus is another addition to the ecology of tools developed in the VIS and HCI community aiming to improve practices in experiment design and statistical analysis.
Like previous works~\cite{Wacharamanotham2015, Eiselmayer2019}, \argus demonstrates the power of direct manipulation interfaces to assist in the tasks previously dominated by menu- or command-based interfaces.
These works add interactivity to existing domain objects (statistical charts and trial tables) to allow the users to \textit{specify}, \textit{compare}, and \textit{explore} diverse outcome possibilities.
These common interaction capabilities and the mappings between abstract concepts in experiment design and statistics to interactive visualizations seems to suggest \textbf{an emerging design pattern for a more usable software tools for research scientists}.

The challenges that these works---including \argus---face is the limited user to participate in evaluation studies.
In other words, our studies have low power--- while we are advocating for the importance of powerful studies.
Specifically, we face \textbf{a trade-off between the coverage of use cases (\eg, which experiment designs to support) and realism of the studies.
}
For \argus, we set the scope of use cases by  pre-determining the scenarios for the study participants.
Although this makes the implementation tractable, the participants might be less motivated to explore---compared to when they design their own experiments.
However, researchers usually design and conduct only a few experiments per year, which imposes a challenge of collecting meaningful longitudinal data.
On the other hand, one could assess learning achievements by novices (\eg, as in~\cite{Wacharamanotham2015}), 
but it is unclear how much the design implications drawn from such learning studies could apply to experts.
In summary, \textbf{we need a methodology that allows studying infrequent knowledge works being conducted by experts}.

%% file: _sections/99_conclusion.tex
%!TEX root=../_main.tex

% --------------------------------------------------------------------
\section{Conclusion}
% --------------------------------------------------------------------

Our goal is to help VIS and HCI researchers consider statistical power when planning their experiments with human participants, which requires performing \textit{a priori} power analysis.
This paper provides four key contributions.
First, we present a detailed \textbf{analysis} of the problems faced by experimenters and identified key challenges and abstract tasks.
% \vspace{-.1cm}

%\begin{enumerate}[nosep]
%    \item Estimating a reasonable effect size is difficult;
%    \item Comparing power across multiple effect size scenarios is necessary; 
%    \item Standardized effect sizes are not intuitive; and
%    \item Incorporating  confounds  into  standardized  effect  size is tricky. 
%\end{enumerate}

Second, we describe the design and implementation of \textbf{\argus}\footnote{Argus is openly available at https://zpac-uzh.github.io/argus/}, an interactive tool for exploring statistical power, and illustrate how it addresses each of the challenges above. 
\argus is the first direct-manipulation tool that lets researchers (1) dynamically explore the relationships among input parameters such as expected averages or potential confounds, statistical outcome, and power;  and (2) evaluate the trade-offs across different experiment design choices.

\changed{%
Third, we describe a \textbf{use case} of designing a visualization experiment based on real studies published in TVCG and CHI. 
The use case illustrates how \argus could be used to incorporate information from prior work and explore possible outcome and power scenarios, resulting in an informed decisions for pilot studies and the actual experiment.
}

Finally, we conducted a \textbf{think-aloud study} to assess how 
\argus helps researchers gain insights from exploring relationships among experiment design concepts and statistical power. % WEM need to say the key findings from the study.
We found that \argus helped both junior and senior researchers to better understand and appreciate the importance of statistical power when conducting controlled experiments. 

We view \argus as a first step towards an ecology of interactive software tools that improve the rigor of designing and conducting experiments in VIS, HCI, and beyond.

% OLD conclusion
% Our goal is to encourage HCI researchers to consider statistical power when planning their experiments.
% Towards this goal, we present three key contributions:
% First, results of \textbf{a preliminary study} reveals that researchers \wm{successfully found} information relevant to power analysis.
% Based on this finding, we present \textbf{\argus}, an interactive simulation tool that lets researchers explore the relationship among experiment design parameters, statistical outcome, and power.
% % Lastly, a \textbf{think-aloud study} shows that researchers could gain insights by exploring the relationship among concepts in experiment design and statistical power through \argus.
% We hope that \argus is a stepping stone towards an ecology of interactive software tools that support works of expert scientists.
% \alex{maybe we can say something stronger here. E.g., We hope that \argus will contribute to the understanding and importance of statistical power when conducting controlled experiments.}